# Computing Optimal Security Strategies for Interdependent Assets

**Joshua Letchford**
Duke University
Durham, NC
jcl@cs.duke.edu

**Yevgeniy Vorobeychik**
Sandia National Laboratories*
Livermore, CA
yvorobe@sandia.gov

## Abstract

We introduce a novel framework for computing optimal randomized security policies in networked domains which extends previous approaches in several ways. First, we extend previous linear programming techniques for Stackelberg security games to incorporate benefits and costs of arbitrary security configurations on individual assets. Second, we offer a principled model of failure cascades that allows us to capture both the direct and indirect value of assets, and extend this model to capture uncertainty about the structure of the interdependency network. Third, we extend the linear programming formulation to account for exogenous (random) failures in addition to targeted attacks. The goal of our work is two-fold. First, we aim to develop techniques for computing optimal security strategies in realistic settings involving interdependent security. To this end, we evaluate the value of our technical contributions in comparison with previous approaches, and show that our approach yields much better defense policies and scales to realistic graphs. Second, our computational framework enables us to attain theoretical insights about security on networks. As an example, we study how allowing security to be endogenous impacts the relative resilience of different network topologies.

## 1  Introduction

Game theoretic approaches to security have received much attention in recent years. There have been numerous attempts to distill various aspects of the problem into a model that could be solved in closed form, particularly accounting for interdependencies of security decisions (e.g., Kunreuther and Heal [2003], Grossklags et al. [2008]). Numerous others offer techniques based on mathematical programming to solve actual instances of security problems. One important such class of problems is network interdiction [Cormican et al., 1998], which models zero-sum encounters between an interdictor, who attempts to destroy a portion of a network, and a smuggler, whose goal typically involves some variant of a network flow problem (e.g., maximizing flow or computing a shortest path).

Our point of departure is another class of optimization-based approaches in security settings: Stackelberg security games [Paruchuri et al., 2008]. These are two-player games in which a *defender* aims to protect a set of targets using a fixed set of limited defense resources, while an *attacker* aims to assail a target that maximizes his expected utility. A central assumption in the literature on Stackelberg security games is that the defender can commit to a probabilistic defense (equivalently, the attacker observes the probabilities with which each target is covered by the defender, but not the actual defense realization).

Much of the work on Stackelberg security games focuses on building fast, scalable algorithms, often in restricted settings [Kiekintveld et al., 2009, Jain et al., 2010, Shieh et al., 2012]. One important such restriction is to assume that targets exhibit *independence*: that is, the defender's utility only depends on which target is attacked and the security configuration at that target. Short of that restriction, one must, in principle, consider all possible combinations of security decisions jointly for all targets, making scalable computation elusive. Many important settings, how-ever, exhibit interdependencies between potential targets of attack. These may be explicit, as in IT and supply chain network security, or implicit, as in defending critical infrastructure (where, for example, successful delivery of transportation services depends on a highly functional energy sector, and vice versa), or in securing complex software systems (with failures at some modules having potential to adversely affect other modules). While in such settings the assumption of independence seems superficially violated, we demonstrate below that under realistic assumptions about the nature of interdependencies, we can nevertheless leverage the highly scalable optimization techniques which assume independence.

In all, we offer the following contributions. (1) We modify and extend the previous linear programming techniques for Stackelberg security games to allow for an arbitrary set of security configurations (rather than merely to cover a target, or not), as well as to account for both random and targeted failures, and replace hard constraints on defense resources with costs associated with specific security configurations (Section 3). (2) We present and justify a crucial assumption on the nature of interdependencies that allows us to use our LP formulations which fundamentally assume independence between targets (Section 4.1). We then offer a simple model of interdependencies based on probabilistic failure cascades satisfying this assumption. Our model makes an explicit distinction between an intrinsic and indirect value of assets, the latter being due entirely to interdependencies. This allows an economically meaningful extension of a well-known independent cascade approach to modeling the spread of infectious diseases or ideas (Section 4). (3) We demonstrate that for trees we can compute expected utilities for all targets in linear time (Section 4). (4) We extend our model to capture uncertainty about network structure, and experimentally study the impact of such uncertainty (Sections 4.5 and 6.1). (5) We show that our approach is both scalable to realistic security settings and offers much better solutions than state-of-the-art alternatives (Sections 5.2 and 5.2). (6) We experimentally study the properties of optimal defense configurations in real and generated networks (Section 6).

## 2 Stackelberg Security Games

A Stackelberg security game consists of two players, the leader (defender) and the follower (attacker), and a set of possible targets. The leader can decide upon a randomized policy of defending the targets, possibly with limited defense resources. The follower (attacker) is assumed to observe the randomized policy of the leader, but not the realized defense actions. Upon observing the leader's strategy, the follower chooses a target so as to maximize its expected utility.

In past work, Stackelberg security game formulations focused on defense policies that were costless, but resource bounded. Specifically, it had been assumed that the defender has $K$ fixed resources available with which to cover targets. Additionally, security decisions amounted to covering a set of targets, or not. While in numerous settings to which such work has been applied (e.g., airport security, federal air marshal scheduling) this formulation is very reasonable, in other settings one may choose among many *security configurations* for each valued asset, and, additionally, security resources are only available at some cost. For example, in cybersecurity, protecting computing nodes could involve configuring anti-virus and/or firewall settings, with stronger settings carrying a benefit of better protection, but at a cost of added inconvenience, lost productivity, as well as possible licensing costs. Indeed, costs on resources may usefully replace resource constraints, since such constraints are often not hard, but rather channel an implicit cost of adding further resources.

While security games as described above naturally entail an attacker, most systems exhibit failures that are not at all a deliberate act of sabotage, but are due entirely to inadvertent errors. Even though such failures are generally far more common than attacks, the vast majority of work in security games posits an attacker, but ignores such failures entirely; essentially the lone exception is a paper by Zhuang and Bier [2007] which offers an analytic treatment of a simple model making explicit the distinction between attacks and natural disasters. Our formulation below is, to our knowledge, the first to explicitly model both attacks and random failures in the Stackelberg security game literature.

To formalize, suppose that the defender can choose from a finite set $O$ of security configurations for each target $t \in T$, with $|T| = n$. A configuration $o \in O$ for target $t \in T$ incurs a cost $c_{o,t}$ to the defender. If the attacker happens to attack $t$ while configuration $o$ is in place, the expected value to the defender is denoted by $U_{o,t}$, while the attacker's value is $V_{o,t}$. A key assumption in Stackelberg security games is that the targets are completely independent: that is, player utilities only depend on the target attacked and its security configuration [Kiekintveld et al., 2009]. We revisit this assumption below when we turn to networked (interdependent) settings. We denote by $q_{o,t}$ the probability that the defender chooses $o$ at target $t$. Finally, let $r$ be the prior probability of the defender that a failure will happen due to a deliberate attack. If no attack is involved, any target can fail; the defender's belief that target $t$ randomly fails (conditional on the event that no attack is involved) is $g_t$, with $\sum_t g_t = 1$.

# 3 Computing Optimal Randomized Security Configurations

Previous formulations of Stackelberg security games involved a fixed collection of defender resources, and in most cases a binary decision to be made for each target: to cover it, or not. To adapt these to our domains of interest, we first modify the well-known multiple linear program (henceforth, multiple-LP) formulation that assumes target independence to incorporate an arbitrary set of security configurations, together with their corresponding costs of deployment. In the multiple-LP formulation, each linear program solves for an optimal randomized defense strategy *given that the attacker attacks a fixed target $\hat{t}$*, with the constraint that $\hat{t}$ is an optimal choice for the attacker. The defender then chooses the best solution from all feasible LPs as his optimal randomized defense configuration. The independence assumption becomes operational here because we treat defense configurations $q_{o,t}$ for each target in isolation, as this assumption obviates the need to randomize over joint defense schedules for all targets. The LP formulation for a representative target $\hat{t}$ is shown in Equations 1a-1d.

$$\max r \left( \sum_o U_{o,\hat{t}} q_{o,\hat{t}}^{\hat{t}} \right) + (1-r) \left( \sum_{t,o} g_t U_{o,t} q_{o,t}^{\hat{t}} \right)$$
$$- \sum_t \sum_o c_{o,t} q_{o,t}^{\hat{t}}. \quad (1a)$$

s.t.

$$\forall_{o,t} \ q_{o,t}^{\hat{t}} \in [0,1] \quad (1b)$$

$$\forall_t \sum_o q_{o,t}^{\hat{t}} = 1 \quad (1c)$$

$$\forall_t \sum_o V_{o,t} q_{o,t}^{\hat{t}} \leq \sum_o V_{o,\hat{t}} q_{o,\hat{t}}^{\hat{t}} \quad (1d)$$

The intuition behind the multiple-LP formulation is that in an optimal defense configuration, the attacker must (weakly) prefer to attack *some* target, and, consequently, one of these LPs must correspond to an optimal defense policy.

Notice that we can easily incorporate additional linear constraints. For example, it is often useful to add a budget constraint of the form:

$$\forall_{\hat{t},t} \quad \sum_o c_{o,t} q_{o,t}^{\hat{t}} \leq B.$$

# 4 Incorporating Network Structure

## 4.1 A General Model of Interdependencies

Thus far, a key assumption has been that the utility of the defender and the attacker for each target depends only on the defense configuration for that target, as well as whether it is attacked or not. In many domains, such as cybersecurity and supply chain security, assets are fundamentally interdependent, with an attack on one target having potential consequences for others. In this section, we show how to transform certain important classes of problems with interdependent assets into a formulation in which targets become effectively independent, for the purposes of our solution techniques.

Below we focus on the defender's utilities; attacker is treated identically. Let $w_t$ be an *intrinsic worth* of a target to the defender, that is, how much loss the defender would suffer if this target were to be compromised with no other target affected (i.e., not accounting for indirect effects). In doing so, we assume that these worths are independent for different targets. Let $s = \{o_1, \ldots, o_n\}$ be the security configuration on all nodes. The probability that a given $t'$ is affected depends on $s$ and the target $t$ chosen by the attacker. Let $z_{s,t'}(t)$ be the marginal probability that target $t'$ is affected when the attacker attacks target $t$. Assuming that the utility function is additive in target-specific worths and the attacker can only attack a single target, the defender's expected utility from choosing $s$ when $t$ is attacked is

$$U_t(s) = E\left[ \sum_{t'} w_{t'} 1(t' \text{ affected} \mid s, t) \right] = \sum_{t'} w_{t'} z_{s,t'}(t),$$

where $1(\cdot)$ is an indicator function. This expression makes apparent that in general $U_t(s)$ depends on defense configurations at all targets, making the problem intractable. We now make the crucial assumption that enables fast computation of defender policies by recovering inter-target independence.

**Assumption 1.** *For all $t$ and $t'$, $z_{s,t'}(t) = z_{o_t,t'}(t)$.*

In words, the probability that a target $t'$ is affected when $t$ is attacked only depends on the security configuration at the attacked target $t$. Below, we use $o$ instead of $o_t$ where $t$ is clear from context.

A way to interpret our assumption is that security against external threats is not very efficacious once an attack has found a way into the system. Alternatively, if the utility of nodes is derived from their contribution to overall connectivity (e.g., in communication networks, where removing a node can, for example, increase latency), it is quite natural to assume that removal of a node impacts global connectivity *regardless of security policies on other nodes*. Our assumption was also operational in other work on interdependent security [Kunreuther and Heal, 2003], where a justification is through a story about airline baggage screening: baggage that is transferred between airlines is rarely thoroughly screened, perhaps due to the expense. Thus, even while an airline may have very strong screening policies, it is poorly protected from luggage entering its planes via transfers. Cybersecurity has similar shortcomings: defense is often focused on external threats, with little attention paid to threats coming from computers internal to the network. Thus, once a computer on a network is compromised, the attacker may find it much easier to compromise others on the same network.

Under the above assumption, the defender utility when $t$ is attacked under security configuration $o$ is:

$$U_{o,t} = z_{o,t}(t)w_t + \sum_{t' \neq t} z_{o,t'}(t)w_{t'}.$$

By a similar argument and an analogous assumption for the attacker's utility, we thereby recover target independence required by the linear programming formulations above.

### 4.2 Cascading Failures Model

In general, one may use an arbitrary model to compute or estimate $z_{o,t'}(t)$. Here, we offer a specific model of interdependence between targets that is simple, natural, and applies across a wide variety of settings.

Suppose that dependencies between targets are represented by a graph $(T, E)$, with $T$ the set of targets (nodes) as above, and $E$ the set of edges $(t, t')$, where an edge from $t$ to $t'$ (or an undirected edge between them) means that target $t'$ depends on target $t$ (and, thus, a successful attack on $t$ may have impact on $t'$). Each target has associated with it a worth, $w_t$ as above, although in this context this worth is incurred only if $t$ is affected (e.g., compromised, broken). The security configuration determines the probability $z_{o,t}(t)$ that target $t$ is affected if the attacker attacks it *directly* and the defense configuration is $o$. We model the interdependencies between the nodes as independent cascade contagion, which has previously been used primarily to model diffusion of product adoption and infectious disease [Kempe et al., 2003, Dodds and Watts, 2005]; Mounzer et al. [2010] is a rare exception (a similar model was also proposed by Tsai et al. [2012], but involves both a defender and an attacker maximizing impact of information diffusion through cascades). The contagion proceeds starting at an attacked node $t$, affecting its network neighbors $t'$ each with probability $p_{t,t'}$; the contagion then spreads from the newly affected nodes $t'$ to their neighbors, and so on. The contagion can only occur once along any network edge, and once a node is affected, it stays affected through the diffusion process. An equivalent way to model this process is to start with the network $(T, E)$ and remove each edge $(t, t')$ with probability $(1 - p_{t,t'})$. The entire connected component of an attacked node is then deemed affected.

### 4.3 Computing Expected Utilities

Given the independent cascade model of interdependencies between targets, we must compute expected utilities, $U_{o,t}$ and $V_{o,t}$, of the defender and the attacker respectively (note that these are expectations only over the cascades, but not the defender's mixed strategy). In general, we can do so by simulating cascades starting at every node $t$ (using breadth-first search), with expected utility of defender/attacker estimated as a sample average over $K$ simulated cascades (expectation in this case is with respect to random realizations of attack success for specific targets as well as edges that become a part of the failure contagion). In several special cases, however, we can either compute these exactly and efficiently, or speed up utility estimation. We now address these special cases.

#### 4.3.1 Cascades on Trees

It is intuitive that when the dependency graph is a tree, expected utilities can be computed efficiently. A naive algorithm can do it in linear time *for each target* $t$, yielding quadratic time in total (since we must repeat the process for all targets). In fact, we can do it in linear time *for all targets*, as the following theorem asserts.

**Theorem 1.** *If $(T, E)$ is an undirected tree we can compute expected utilities for at targets in $O(|T|)$ time.*

The proofs of this and other results can be found in the online supplement.

#### 4.3.2 Cascades on Undirected Graphs

In general undirected graphs, we can apply a very simple optimization in the way we sample cascades to obtain substantial speedups when the graph is dense. First, observe that rather than determining live edges as the cascade unfolds, we can instead flip the biased coin for each edge to determine whether it is live or not during a particular cascade *prior* to propagating the failure. The resulting graph contains a subset of edges from the original graph. At this point, observe that each potential target in a given connected component will result in the same defender/attacker utility. We therefore only need to compute the expected loss once for each connected component. When the size of the largest connected component is $O(|T|)$, a likely scenario in dense graphs, this optimization results in an $O(|T|)$ speedup.

### 4.4 The Significance of Capturing Interdependence

An obvious question that may arise upon pondering the complexities of our framework is whether they are worthwhile: it may well be that previous approaches which assume target independence offer satisfactory approximation. We now show theoretically, and later experimentally, that our approach improves dramatically as compared to one which assumes independence.

**Proposition 1.** *There exists a family of problem instances for which the independence assumption yields a solution that is a factor of $O(n)$ worse than optimal.*

### 4.5 Incorporating Uncertainty about the Network

Applying our framework in real-world networked security settings requires an accurate understanding of the interdependencies. Thus far, we assumed that the actual network over which cascading failures would spread is perfectly known. A natural question is: what if our network model is inaccurate?

Formally, we model the uncertainty about the network as a parameter $\epsilon$ which represents the probability of incorrectly estimating the relationship between a pair of targets. Thus, if there is an edge between $t$ and $t'$, we now let this edge be present with probability $1 - \epsilon$. On the other hand, if $t$ and $t'$ are not connected in the graph given to us, we propose that they are, in fact, connected with probability $\epsilon$. Thus, when the graph is large, even a small amount noise will cause us to err about a substantial number of edges.[1]

Note that there is a natural way to incorporate this model of uncertainty into our framework. Let us interpret $p_{t,t'}$ as the probability of a cascade from $t$ to $t'$ *conditional on an edge from $t$ to $t'$*. Then, if $t$ and $t'$ are connected, we modify cascade probabilities to be $\hat{p}_{t,t'} = p_{t,t'}(1 - \epsilon)$, whereas if they are not connected, the cascade probability is $\hat{p}_{t,t'} = p_{t,t'}\epsilon$.

## 5 Experiments

The goal of this section is to illustrate the value of our framework as a computational tool for designing security in interdependent settings. Specifically, we aim to demonstrate that our approach clearly improves on state-of-the-art alternatives, and offers a scalable solution for realistic security problems. We pursue this aim by randomly constructing dependency graphs using Erdos-Renyi (ER) and Preferential Attachment (PA) generative models [Newman, 2010], as well as using a graph representing a snapshot of Autonomous System (AS) interconnections generated using Oregon routeviews [of Oregon Route Views Project]; this graph contains 6474 targets and 13233 edges and thus offers a reasonable test of scalability. In the ER model every directed link is made with a specified and fixed probability $p$; we refer to it as $ER(p)$. The PA model adds nodes in a fixed sequence, starting from an arbitrary seed graph with at least two vertices. Each node $i$ is attached to $m$ others stochastically (unless $i \leq m$, in which case it is connected to all preceding nodes), with probability of connecting to a node $j$ proportional to the degree of $j$, $d_j$.

For the randomly generated networks, all data presented is averaged over 100 graph samples. Since we generate graphs that may include undirected cycles, we obtain expected utilities for all nodes on a given graph using 10,000 simulated cascades (below we show that this is more than sufficient). Intrinsic worths $w_t$ are generated uniformly randomly on $[0, 1]$. Cascade probabilities $p_{t,t'}$ were set to 0.5 unless otherwise specified. In the sequel, we restrict the defender to two security configurations at every target, one with a cost of 0 which stops attacks with probability 0 and one with a cost of $c$ which prevents attacks with probability 1.

### 5.1 Sampling Efficiency

Throughout our experiments we use 10,000 samples to evaluate the expected utilities of players. A natural question is: are we taking enough samples? To answer this, we systematically varied the number of samples between 0 (i.e., letting $U_{o,t} = -w_t$) and 100,000. Our results offer strong evidence that 10,000 samples is more than enough: the expected utility (evaluated using 100,000 samples) of the resulting defense configurations becomes flat already when the number of samples is 1000.

### 5.2 Scalability

An important question given the complexity of our framework is whether it can scale to realistic defense scenarios. To test this, we ran our framework on the AS graph consisting of 6474 targets and 13233 edges. Since this is a large undirected graph containing cycles, a sampling approach was required, but the total running time (including both sampling and solving linear programs) amounted to less than 1 hour. Given the importance of security, and the fact that *distributions* of security settings are computed once (or at least infrequently, as long as significant changes to the interdependency structure are not very frequent), this seems a relatively small computational burden.

---
[1] We assume here that both the defender and attacker share the same uncertainty about the network. An alternative model could consider an attacker that has more (or exact) information about the network. The resulting defender problem would become a Bayesian Stackelberg game.

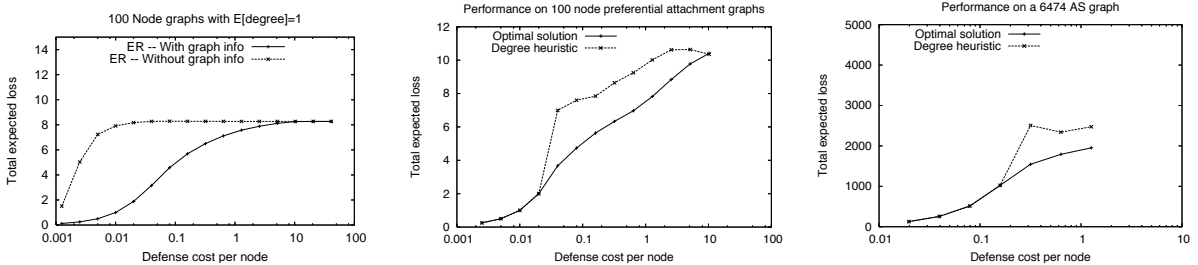

Figure 1: Left: Comparison between our approach ("with graph info") and one assuming independence ("without graph info") using the ER(0.1) generative model. Middle: Comparison to the degree-based heuristic on PA graphs. Right: Comparison to the degree-based heuristic on the AS graph.

### 5.3 Comparison to State-of-the-Art Alternatives

There are two prime computational alternatives to our framework. The first is to assume that targets are independent. While we showed above that in the worst case this can be quite a poor approximation, we offer empirical support to the added value of our approach below. The second is to use a well-known heuristic developed in the context of vaccination strategies on networks. This latter heuristic would in our case defend nodes in order of their connectivity (degree), until the defense budget is exhausted. Figure 1 compares our approach first to the former (left) and then to the latter (right). In both cases, computing optimal defense strategies using our framework yields much higher utility to the defender than the alternatives.

Aside from interdependencies, two other important aspects of our model are the fact that it allows an arbitrary number of security configurations, instead of simply allowing the defender to defend, or not, each target, and its ability to optimize with respect to both intelligent attackers and inadvertent failures. We now show that both of these can add substantial value. Figure 2 (top) shows a comparison between a solution which only allows two configurations (defend and do not defend) and two solutions which also allow for a third configuration, which is less effective than full defense, but also less costly. We consider two potential third options, one providing 50% defense at 12.5% of the cost of full defense ($1/2 - 1/8$) and one providing 75% defense at 12.5% cost ($3/4 - 1/8$). It is clear from this graph that considering the third configuration adds considerable value. Figure 2 (bottom) assumes that all (or nearly all) failures arise randomly, and compares a solution which posits an attacker to an optimal solution. Again, the value of solving the problem optimally is clear. This plot actually shows an interesting pattern, as the expected utility of the defender is non-monotonic in cost when the solution is suboptimal. This is because the differences between the two solutions are most important when costs are intermediate; with low costs, nearly everything is fully defended, while high costs imply almost no defense.

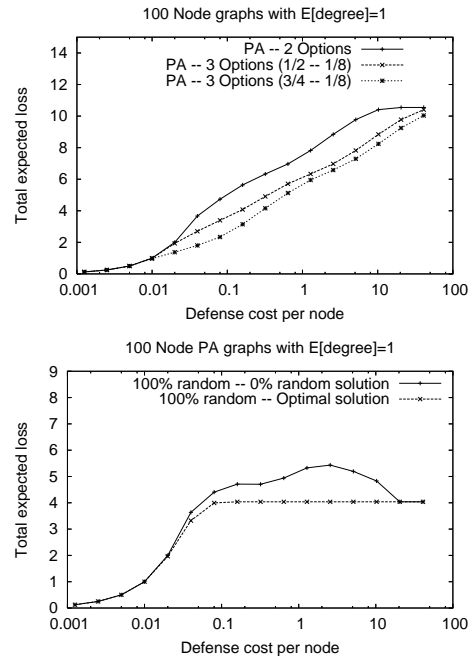

Figure 2: Top: Comparison between assuming only two configurations, and allowing the defender to consider three alternatives. Bottom: Comparison between a solution which assumes that failures are only due to attacks, and an optimal solution, when failures are actually random. Comparisons use PA graphs.

## 6 Applications to Interdependent Security Analysis

In this section we apply our framework to several network security domains. For simplicity, we restrict at-tention to zero-sum security games, unless otherwise specified. As above, we consider ER and PA generative models, although we utilize a generalized version of PA. In a generalized PA model, connection probabilities are $\frac{(d_i)^\mu}{\sum_j (d_j)^\mu}$, such that when $\mu = 0$ the degree distribution is relatively homogeneous, just as in ER, $\mu = 1$ recovers the "standard" PA model, and large values of $\mu$ correspond to highly inhomogeneous degree distributions. Throughout, we use $\mu = 1$ unless otherwise specified. All parameters are set as in the experiments section, unless otherwise specified.

In addition to generative models of networks, we explore two networks derived from real security settings: one with 18 nodes that models dependencies among critical infrastructure and key resource sectors (CIKR), as inferred from the DHS and FEMA websites, and the second with 66 nodes that captures payments between banks in the core of the Fedwire network [Soramaki et al., 2007].

For the CIKR network, each node was assigned a low, medium, or high worth of 0.2, 0.5, or 1, respectively, based on perceived importance (for example, the energy sector was assigned a high worth, while the national monuments and icons sector a low worth). Each edge was categorized based on the importance of the dependency (gleaned from the DHS and FEMA websites) as "highly" or "moderately" significant, with cascade probabilities of 0.5 or 0.1 respectively. For the Fedwire network, all nodes were assigned an equal worth of 0.5, and cascade probabilities were discretely chosen between 0.05 and 0.5 in 0.05 increments depending on the weight of the corresponding edges shown in Soramaki et al. [2007].

### 6.1 The Impact of Uncertainty

Our framework offers a natural way to incorporate uncertainty about the network into the analysis. An important question is: how much impact on defender decision does uncertainty about the network have? Figure 3 quantifies the impact of uncertainty on the quality of defense if the observed graph is the PA network with average degree of 2. When cascade probabilities are relatively high ($p_{t,t'} = 0.5$ for all edges, top plot), even if the amount of noise is relatively small ($\epsilon = 0.01$), the resulting increase in the number of possible cascade paths in the network makes the defender much more vulnerable. With smaller cascade probabilities ($p_{t,t'} = 0.1$, bottom plot), however, noise has relatively little impact. It can thus be vital for the defender to obtain an accurate portrait of the true network over which failures may cascade when the interdependencies among the components are strong.

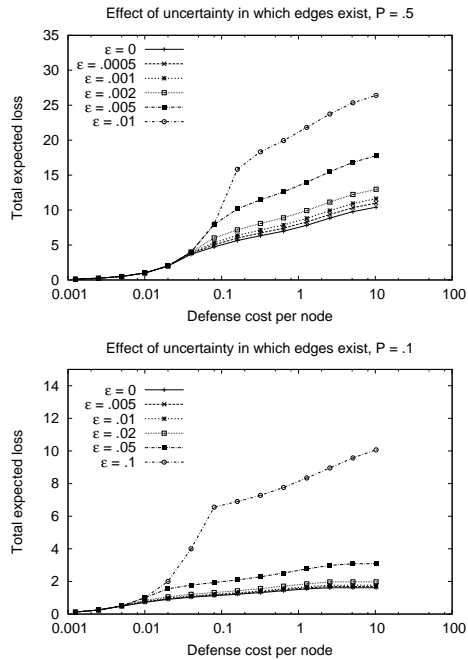

Figure 3: The impact of noise on PA networks. Top: when $p_{t,t'} = 0.5$; Bottom: when $p_{t,t'} = 0.1$.

### 6.2 The Impact of Marginal Defense Cost

Our next analysis deals with the impact of marginal defense cost $c$ on defender expected losses, his total costs, and the sum of these (i.e., negative expected utility). The results for ER and BA (both with 100 nodes and average degree of 2), as well as CIKR and Fedwire networks are shown in Figure 4. All the plots feature a clear pattern: expected loss and (negative) utility are monotonically increasing, as expected, while total costs start at zero, initially rise, and ultimately fall (back to zero in 3 of the 4 cases). It may at first be surprising that total costs eventually fall even as marginal costs continue to increase, but this clearly must be the case: when $c$ is high enough, the defender will not wish to invest in security at all, and total costs will be zero. What is much more surprising is the presence of two peaks in PA and Fedwire networks. Both of these networks share the property that there is a non-negligible fraction of nodes with very high connectivity [Newman, 2010, Soramaki et al., 2007]. When the initial peak is reached, the network is fully defended, and as marginal costs rise further, the defender begins to reduce the defense resources expended on the less important targets. At a certain point, only the most connected targets are protected, and since these are so vital to protect, total costs begin increasing again. After the second peak is reached, $c$ is finally large enough to discourage the defender from fully protecting even the most important targets, and the subsequent fall of total costs is no longer reversed.

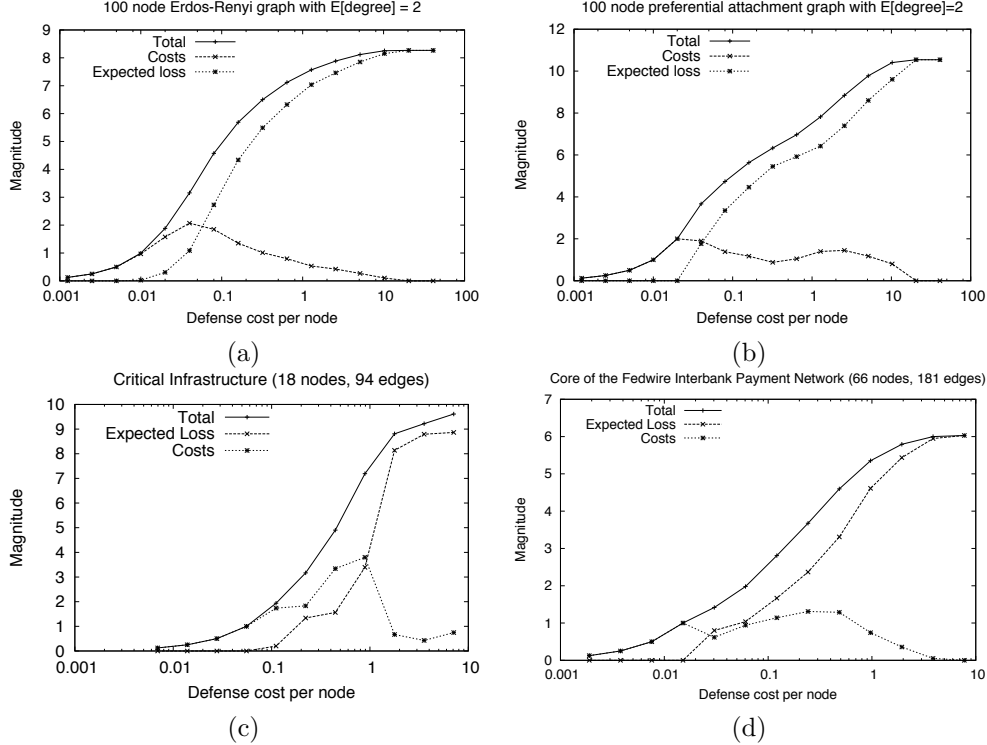

Figure 4: Expected loss, cost, and their sum in (a) 100-node ER(0.2), (b) 100-node PA, (c) 18-node critical infrastructure, and (d) 66-node core of the Fedwire networks as defense cost increases. The results for ER and PA are averages over 100 stochastic realizations of these networks.

### 6.3 Resilience to Targeted Attacks: Impact of Network Structure

One of the important streams in the network science literature is the question of relative resilience of different network topologies to failures, random or targeted. A central result, replicated in a number of contexts, is that network topology is a vital factor in determining resilience [Albert et al., 2000, Newman, 2010]. Of particular interest to us is the observation that scale-free networks such as PA exhibit poor tolerance to targeted attacks as compared to ER [Albert et al., 2000], which is precisely the context that we consider.

In Figure 5 (top) we show the defender's utility for three different network topologies, PA, ER, and Fedwire as a function of cost $c$. Remarkably, there is essentially no difference between PA and ER (and not much between these and Fedwire) until $c$ is quite high, at which point they begin to diverge. This seems to contradict essentially all the previous findings in that network topology seems to play little role in resilience in our case! A superficial difference here is that we consider a cascading failure model, while most of the previous work on the subject focused on diminished connectivity due to attacks. We contend that the most important distinction, however, is that previous work studying resilience did not account for a simple observation that most important targets are also most heavily defended; indeed, there was no notion of endogenous defense at all. In scale-free graphs, there are well connected nodes whose failure has global consequences. These are the nodes which are most important, and are heavily defended in optimal decisions prescribed by our framework. Once the defense decision becomes endogenous, differences in network topology disappear. Naturally, once $c$ is high enough, defense of important targets weakens, and eventually we recover the standard result: for high $c$, PA is considerably more vulnerable than ER.

To investigate the impact of network topology on resilience further, we consider the generalized PA model in which we systematically vary the homogeneity of the degree distribution by way of the parameter $\mu$. The results are shown in Figure 5 (bottom). In this graph, we do observe clear variation in resilience as a function of network topology, but the operational factor in this variation is homogeneity in the distribution of expected utilities, rather than degrees: increasing *homogeneity* of the utility distribution *lowers network resilience*. This seems precisely the opposite of the standard results in network resilience, but the two are in fact closely related, as we now demonstrate. Superficially, the trend in the figure seems to follow the common intuition in the resilience literature: as the degree distribution becomes more inhomogeneous (more star-like), it becomes more difficult to defend. Observe, however, that ER is actually more difficult to defend than PA with $\mu = 0$. The lone difference of the latter from ER is the fact that nodes that enter earlier are more connected and, therefore, the degree distribution in the PA variant should actually be more *inhomogeneous* than ER! The answer is that random connectivity combined with inhomogeneity of degrees actually makes the distribution of *utilities* less homogeneous in PA with $\mu = 0$, and, as a result, fewer nodes on which defense can focus as compared to ER. On the other hand, as the graph becomes more star-like, the utilities of all nodes become quite similar; in the limiting case, all nodes are only two hops apart, and attacking any one of them yields a loss of many as a result of cascades.

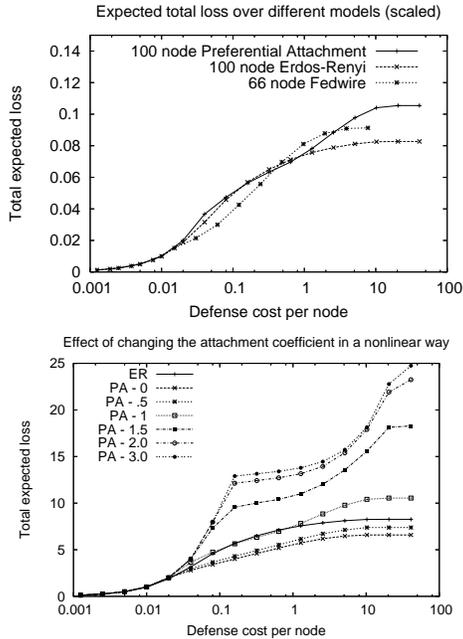

Figure 5: Top: Expected total loss: comparison across different network structures. Bottom: Expected defender disutility in the generalized PA model as we vary $\mu$ (keeping average degree fixed at 2). ER is also shown for comparison.

There is another aspect of network topology that has an important impact on resilience: network density. Figure 6 shows a plot of an Erdos-Renyi network with the probability of an edge varying between 0.0025 to 0.08 (average degree between .25 and 8) and cost $c$ fixed at 0.04. Clearly, expected utility and loss of the defender are increasing in density, but it is rather sur-prising to observe how sharply they jump once average degree exceeds 1 (the ER network threshold for a large connected component); in any case, network density has an unmistakable impact. The reason is intuitive: increased density means more paths between targets, and, consequently, greater likelihood of large cascades in the event that a target is compromised. Total cost initially increases in response to increased density, in part to compensate for the increased vulnerability to attacks, but eventually falls, since it is too expensive to protect everything, and anything short of that is largely ineffective.

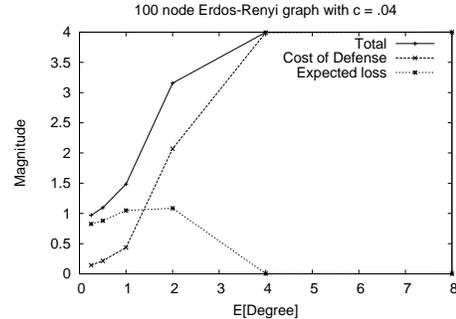

Figure 6: Expected loss, cost, and their sum in 100-node Erdos-Renyi networks as a function of network density (equivalently, expected degree).

## 7 Conclusion

We presented a framework for computing optimal randomized security policies in network domains, extending previous linear programming approaches to Stackelberg security games in several ways. First, we extended previous linear programming techniques to incorporate benefits and costs of arbitrary security configurations on individual assets. Second, we offered a principled model of failure cascades that allows us to capture both the direct and indirect value of assets, and showed how to extend this model to capture uncertainty about the structure of the interdependency network. Third, we allowed the defender to account for failures due to actual attacks, as well as those that are a result of exogenous failures. Our results demonstrate the value of our approach as compared to alternatives, and show that it is scalable to realistic security settings. Furthermore, we used our framework to analyze four models of interdependencies: two based on random graph generation models, a simple model of interdependence between critical infrastructure and key resource sectors, and a model of the Fedwire interbank payment network.